\documentclass[conference]{IEEEtran}
\IEEEoverridecommandlockouts

\usepackage{cite}
\usepackage{amsmath,amssymb,amsfonts}
\usepackage{algorithmic}
\usepackage{graphicx}
\usepackage{multirow}
\usepackage{textcomp}
\usepackage{xcolor}
\def\BibTeX{{\rm B\kern-.05em{\sc i\kern-.025em b}\kern-.08em
    T\kern-.1667em\lower.7ex\hbox{E}\kern-.125emX}}
    
\begin{document}
\fontsize{9pt}{11pt}\selectfont

\title{SLIDE: Integrating Speech Language Model with LLM for Spontaneous Spoken Dialogue Generation\\
\thanks{\textsuperscript{†}Corresponding author.}
\thanks{This work is partially supported by the National Natural Science Foundation of China (No.62401560), the Youth Innovation Promotion Association Chinese Academy of Sciences, the Basic and Frontier Exploration Project Independently Deployed by Institute of Acoustics, Chinese Academy of Sciences (No.JCQY202411).}
}


\author{
	\IEEEauthorblockN{Haitian Lu$^{1,2}$, Gaofeng Cheng$^{1,2,\textit{\dag}}$, Liuping Luo$^{3}$, Leying Zhang$^{4}$, Yanmin Qian$^{4}$, Pengyuan Zhang$^{1,2}$}
	\IEEEauthorblockA{
		$^1$ Key Laboratory of Speech Acoustics and Content Understanding, Institute of Acoustics, CAS, China \\
		$^2$ University of Chinese Academy of Sciences, China \\
            $^3$ Guangdong Provincial Public Security Department, Guangzhou, China \\
            $^4$ AudioCC Lab, MoE Key Lab of AI, CS Dept, Shanghai Jiao Tong University, Shanghai, China \\
        \IEEEauthorblockA{\small \{luhaitian, chenggaofeng, zhangpengyuan\}@hccl.ioa.ac.cn, llp1632024@163.com, \{zhangleying, yanminqian\}@sjtu.edu.cn \\
	}
	}
        }

\maketitle

\begin{abstract}
Recently, ``textless" speech language models (SLMs) based on speech units have made huge progress in generating naturalistic speech, including non-verbal vocalizations. However, the generated speech samples often lack semantic coherence. In this paper, we propose \textbf{S}LM and \textbf{L}LM \textbf{I}ntegration for spontaneous spoken \textbf{D}ialogue g\textbf{E}neration (SLIDE). Specifically, we first utilize an LLM to generate the textual content of spoken dialogue. Next, we convert the textual dialogues into phoneme sequences and use a two-tower transformer-based duration predictor to predict the duration of each phoneme. Finally, an SLM conditioned on the spoken phoneme sequences is used to vocalize the textual dialogue. Experimental results on the Fisher dataset demonstrate that our system can generate naturalistic spoken dialogue while maintaining high semantic coherence.
\end{abstract}


\section{Introduction}
Recent advances in spoken language models (SLMs) have highlighted their superiority in generating naturalistic spoken dialogues, characterized by fluid turn-taking and non-verbal vocalizations \cite{dgslm}. However, SLMs often struggle to maintain semantic coherence and typically require large training datasets \cite{borsos2023audiolm, rubenstein2023audiopalm}. Spoken dialogue encompasses two critical aspects: the semantic and the naturalistic. The semantic aspect pertains to the meaningfulness of the dialogue's content, which is vital to convey accurate and relevant information \cite{ni2023recent}. The naturalistic aspect involves the fluidity of turn-taking, including inter-pausal units (IPUs), overlaps, gaps, pauses, and other naturalistic dialogue events such as laughter and backchannels \cite{ten2005temporal, heldner2010pauses, shortphrase}. This aspect is essential for creating dialogue that feels authentic and engaging, sustaining the listener's interest, and ensuring that the interaction appears genuine and relatable.

Traditional spoken dialogue generation systems are typically cascaded \cite{cascadesqa, tseng2016towards}. These systems encompass several components: automatic speech recognition (ASR) for transcribing spoken input into text \cite{10737652}, LLMs for textual dialogue generation, and text-to-speech (TTS) for synthesizing spoken dialogue from the generated text. Although these systems exhibit strong semantic coherence, thanks to LLMs trained on datasets containing tens of billion words, their ability to produce naturalistic dialogues is still very limited \cite{traum2012ada}. This limitation arises because they do not account for turn-taking events within any of the components \cite{SKANTZE2021101178}. Additionally, these systems struggle to generate natural dialogues that include laughter and backchannels \cite{inoue2022can}, as encoding speech as text during the intermediate stage results in the loss of paralinguistic information, such as non-verbal vocalizations.

\begin{figure*}[htb]
    \centering
    \includegraphics[width=\textwidth]{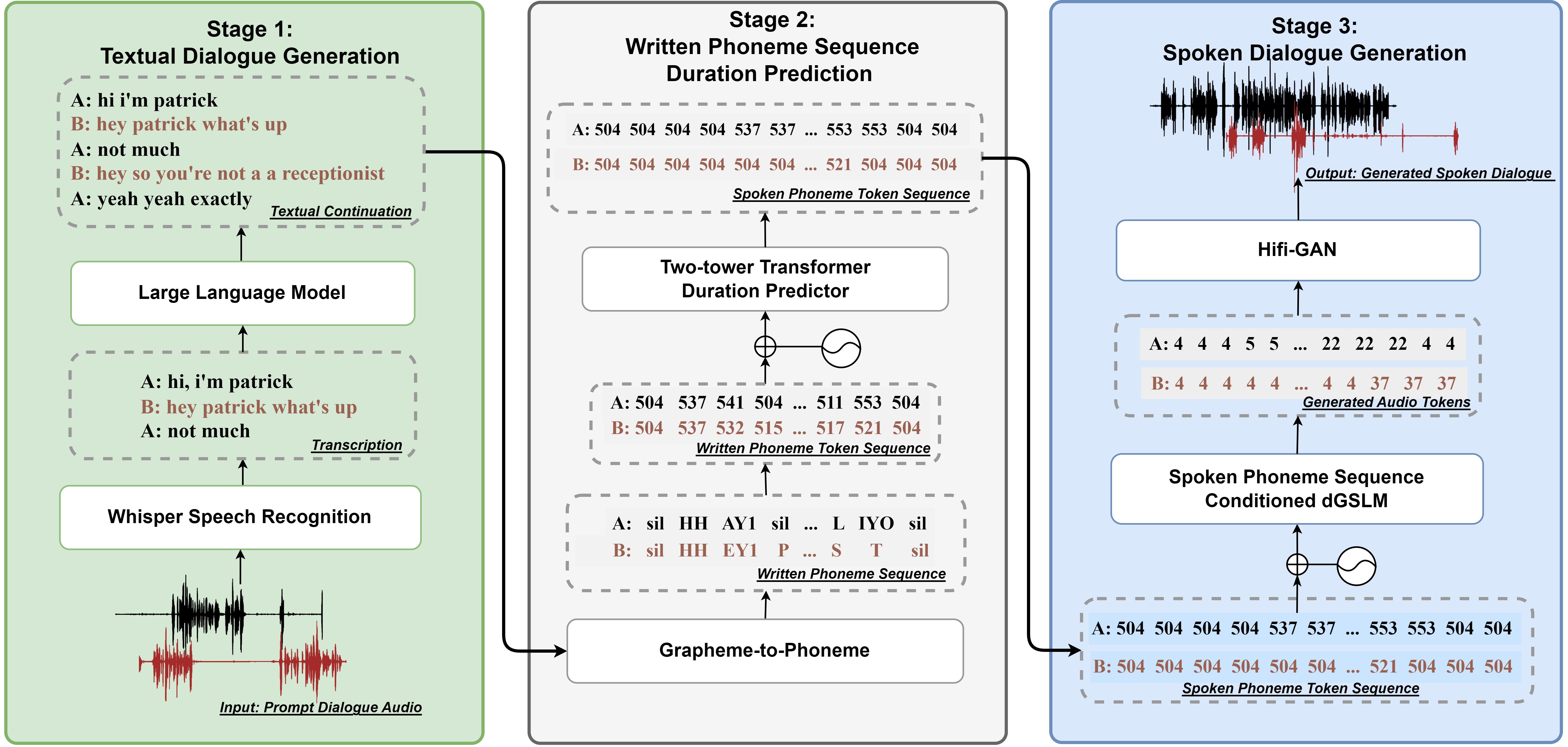}
    \caption{The inference diagram of the proposed SLIDE model for spoken dialogue generation, with black representing elements from Channel A and brown representing elements from Channel B. Written phoneme sequence refers to phonemes obtained from grapheme-to-phoneme (G2P) conversion of textual dialogues, with silent phonemes inserted between different sentences. Spoken phoneme sequence extends the phonemes from the written phoneme sequence by repeating them to represent their durations in speech.}
    \label{fig:wide_pdf_picture}
\end{figure*}

Another approach to spoken dialogue generation leverages speech language models (SLMs) \cite{gslm,zhang2024covomix,mitsui2023towards}, with the generative spoken dialogue language model (dGSLM) \cite{dgslm} being a notable example. Unlike traditional cascaded systems, dGSLM replaces text-based encoding units with discrete speech units derived from self-supervised models \cite{hsu2021hubert, baevski2020wav2vec, ye2023asq, gao2022self}. It comprises three key components: a HuBERT-based speech-to-units encoder \cite{hsu2021hubert}, a two-tower transformer-based unit language model (uLM) \cite{vaswani2017attention}, and a Hifi-GAN-based units-to-speech decoder \cite{HifiGAN, kong2020hifi}. Because dGSLM operates entirely on speech units, it can effectively capture dialogue elements such as laughter and backchannels \cite{liu2024audioldm}, which are often absent in text transcriptions. Additionally, since these speech units are time-aligned with raw audio, they inherently encode the turn-taking patterns of spoken dialogue, allowing the uLM to learn and reproduce fluid turn-taking dynamics. However, dGSLM faces challenges in generating semantically coherent spoken dialogue. The speech units are fine-grained, typically spanning only 20 ms \cite{borsos2023audiolm, rubenstein2023audiopalm}, which makes them unsuitable for modeling semantic content over extended contexts. Furthermore, the fine granularity of these speech units significantly increases the need for large training datasets \cite{chang2024exploring}.

In this paper, we propose the SLIDE model, a spontaneous spoken dialogue generation model that employs text to capture semantic context, while using speech units to preserve paralinguistic information, such as non-verbal vocalizations and turn-taking patterns\footnote{Generation samples can be found at https://github.com/SpeechClub/SLIDE\\-demo/tree/main}. The contributions of this paper can be summarized as follows:%

1. Incorporating LLMs into the spoken dialogue generation framework to generate textual dialogues, thus leveraging the advanced text generation capabilities of LLMs.

2. Utilizing a two-tower transformer-based model to predict the duration of each phoneme in the written phoneme sequence, ensuring the preservation of turn-taking fluidity.

3. Conditioning dGSLM on spoken phoneme sequences derived from textual dialogues, effectively incorporating natural dialogue events into the generated speech while maintaining semantic coherence.

The rest of this paper is organized as follows: In Section \ref{sec:method}, we describe the proposed SLIDE model. The experiments and conclusions are presented in Sections \ref{sec:exp} and \ref{sec:conclusion}, respectively.


\section{The Proposed SLIDE Model}
\label{sec:method}

As depicted in Fig. \ref{fig:wide_pdf_picture}, the proposed method consists of three main parts: textual dialogue generation, written phoneme sequence duartion prediction, and spoken dialogue generation. 

For textual dialogue generation, we utilize LLMs to produce verbal backchannels such as ``yeah," ``right," and ``okay," as well as brief comments that signal a listener's attentiveness. The generation of non-verbal vocalizations is addressed by dGSLM in Section \ref{ssec:subhead3}.

For written phoneme sequence duration prediction, we propose this method because simply conditioning dGSLM on written phoneme sequences leads to significant errors in the generated spoken dialogue, including the omission of multiple utterances present in the textual dialogues.

For spoken phoneme sequence-conditioned dGSLM, we apply spoken phoneme sequences as a conditioning mechanism because the original dGSLM struggles with generating semantically coherent dialogues. Our proposed method constrains the output speech units of dGSLM with the corresponding phonemes, ensuring that the generated spoken dialogues maintain semantic coherence.

\subsection{LLM-driven Textual Dialogue Generation}
\label{ssec:subhead1}

As shown in the left block of Fig. \ref{fig:wide_pdf_picture}, we utilize LLMs to generate the textual dialogue continuations.

Firstly, we use speech recognition models to transcribe the prompt dialogue audio into text. Secondly, we utilize LLMs to generate the continuation of the textual dialogue. We instruct LLMs to produce dialogue continuations in the style of spoken dialogues. Specifically, we exclude dialogue event markers such as [laughter] and [sigh] from the ground-truth transcriptions in the training dataset. The LLM is provided with a sample spoken dialogue transcription, along with the following instruction: ``Generate conversations in a similar style but on different topics. The conversations should maintain a casual, conversational tone with ample use of fillers and backchannels. Short utterances may consist of single words like `yes,' `okay,' `right,' and so on. Each conversation should include at least 20 turns of dialogue."

\subsection{Two-tower Transformer-based Written Phoneme Sequence Duration Prediction}
\label{ssec:subhead2}


As shown in the middle block of Fig. \ref{fig:wide_pdf_picture}, we employ two-tower transformers proposed in \cite{dgslm} to predict the duration of each phoneme in the written phoneme sequences.

We used forced alignment models to obtain phoneme-level text-to-speech alignments for the ground-truth transcriptions in the training dataset. The alignment results are then employed to generate spoken phoneme sequences. To enhance the phoneme set, we introduce an additional silence phoneme. Each phoneme, including the silence phoneme, is repeated multiple times according to its duration as determined by the forced alignment. Specifically, we divide the duration of each phoneme and silence segment by 20 ms, and the resulting value is rounded to determine the number of repetitions for each phoneme in the spoken phoneme sequences.

During the training phase, the spoken phoneme sequences derived from forced alignment results serve as both the input and the target for the two-tower transformer duration predictor. Training is conducted using a teacher-forcing approach. The loss function used in training is a combination of the edge unit loss and the edge duration loss. We also use the delayed duration prediction as detailed in \cite{dgslm}.

During the inference phase, we perform an unconditional generation. If the phoneme generated in the current time step differs from the penultimate generated phoneme, we replace it with the corresponding phoneme from the written phoneme sequence. For subsequent replacements, we continue to use the next phoneme in the written phoneme sequence.

To prevent excessive overlap in the generated spoken phoneme sequences, we introduce a post-processing step that inserts silence tokens before the speech region whenever an overlap exceeds 0.6 seconds, ensuring that the overlap does not exceed 0.3 seconds.

\subsection{Spoken Phoneme Sequence Conditioned dGSLM for Spoken Dialogue Generation}
\label{ssec:subhead3}

As shown in the right block of Fig. \ref{fig:wide_pdf_picture}, we condition the dGSLM model on spoken phoneme sequences and generate corresponding audio token continuations.

During the training phase, we use a Hubert-based speech-to-unit model to encode spoken dialogues into audio tokens. The concatenated spoken phoneme sequence and audio tokens serve as both the input and training target. Each dialogue audio sample is segmented into 80-second intervals. Both the spoken phoneme sequences and audio tokens have a 20 ms granularity, resulting in 8,000 discrete tokens per sample. The first 4,000 tokens represent spoken phoneme sequences, while the remaining 4,000 tokens correspond to audio tokens. The training process follows the scheme proposed in \cite{dgslm} for dGSLM. Distinct positional encodings are applied to spoken phoneme sequences and the audio tokens, allowing the model to learn the correspondence between them. 

During the inference phase, we adjust the spoken phoneme sequence to a fixed length of 4,000 tokens by either cropping or padding, and use it as the model input. We then continue to generate audio tokens autoregressively until the token sequence length exceeds 8,000. Finally, the last 4,000 generated tokens are fed into a HiFi-GAN-based units-to-speech model to decode the spoken dialogue. Each generated dialogue corresponds to an 80-second duration.

\section{Experiments}
\label{sec:exp}

\subsection{Implemention Details}

All experiments are conducted using the Fisher dataset \cite{cieri2004fisher}, which consists of 2000 hours of stereo telephone conversation audio sampled at 8 kHz. We resample the audio to 16 kHz using torchaudio \cite{hwang2023torchaudio} to follow the experiment setup in \cite{dgslm}.

For transcribing the prompt dialogue audio into text, we employ the base Whisper-v3 speech recognition model \cite{radford2023robust}. After transcribing each channel of the stereo audio, we insert the channel ID at the beginning of each utterance and rearrange them according to the timestamps provided by the Whisper model. To generate textual dialogue continuations, we utilize GPT-4o \cite{achiam2023gpt}, one of the best-performing large language models (LLMs). For converting textual dialogue into phoneme sequences, we use the G2P toolkit \cite{G2PE2019}, and for forced alignment, we employ the Montreal Forced Aligner (MFA) \cite{mcauliffe2017montreal}.

Regarding the model configuration of the two-tower transformer-based phoneme sequence duration predictor and the spoken phoneme sequence-conditioned dGSLM, we follow the setup outlined in \cite{dgslm}. The Hubert model and the discrete unit-based Hifi-GAN model are identical to those used in \cite{dgslm}. For the two-tower transformer model used for written phoneme duration prediction, we train the model on the 2000-hour stereo audio data from the Fisher dataset, with each training sample containing up to 4,000 unit pairs. The model is trained on six A100 40GB GPUs, with a batch size of 48 for 50000 steps. Similarly, the two-tower transformer model used for spoken phoneme sequence conditioned dGSLM is trained on the same 2000-hour stereo audio, with each sample containing up to 8000 unit pairs. This model is also trained on six A100 40GB GPUs, with a batch size of 96 for 250000 steps.

\begin{table}[b]%
    \centering
    \renewcommand\arraystretch{1.1}
    \caption{Number of Turn-taking Events per Minute}
    \fontsize{9pt}{11pt}\selectfont
    \setlength{\tabcolsep}{4mm}{
    \begin{tabular}{l|cccc}
        \hline
        \multirow{2}{*}{Model}& \multicolumn{4}{c}{Number of occurrence / min}\\
        & IPU & Pause & Gap & Overlap \\
        \hline\hline
        Cascaded \cite{dgslm} & 17.5 & 0.0 & 14.9 & 0.0 \\
        dGSLM & 30.6 & 12.0 & 9.0 & 8.7 \\
        SLIDE-1 & 25.6 & 9.4 & 5.6 & 9.5 \\
        SLIDE-2 & 31.3 & 6.3 & 7.6 & 15.8 \\
        \hline
        Ground Truth & 27.3 & 9.9 & 8.9 & 8.2 \\
        \hline
    \end{tabular}%
    }
    
    \label{tab:occurrences_per_min}
\end{table}

\begin{table}[b]%
    \centering
    \renewcommand\arraystretch{1.1}
    \caption{Cumulated Durations of Turn-taking Events per Minute}
    \fontsize{9pt}{11pt}\selectfont
    \setlength{\tabcolsep}{4mm}{
    \begin{tabular}{l|cccc}
        \hline
        \multirow{2}{*}{Model}& \multicolumn{4}{c}{Cumulated durations / min}\\
         & IPU & Pause & Gap & Overlap \\
        \hline\hline
        Cascaded \cite{dgslm} & 54.8s & 0.0s & 5.3s & 0.0s \\
        dGSLM & 53.5s & 7.1s & 4.0s & 4.9s \\
        SLIDE-1 & 54.3s & 5.6s & 2.4s & 3.6s \\
        SLIDE-2 & 57.8s & 3.5s & 3.6s & 7.0s \\
        \hline
        Ground Truth & 54.5s & 5.8s & 3.7s & 4.0s \\
        \hline
    \end{tabular}%
    }
    
    \label{tab:durations_per_min}
\end{table}

\subsection{Evaluation Metrics}
\label{subsec: metric}

We consider both subjective and objective evaluation metrics. For subjective evaluation, participants are asked to assess two key aspects of the spoken dialogue: naturalness and semantic coherence. Regarding naturalness, participants focus on (1) the presence of natural dialogue events, such as laughter and backchannels, and (2) the fluidity of turn-taking in the dialogue. For semantic coherence, participants evaluate whether the spoken dialogue maintains logical consistency and meaning. Each audio sample is rated on a scale from 1 to 5, with 1 being the worst and 5 being the best. Each audio sample is rated by at least 5 participants.

For objective evaluations, we also focus on two aspects of spoken dialogue: naturalism and semantic coherence. To assess naturalism, we compare the temporal distribution of turn-taking events in the generated spoken dialogue with that in the ground-truth spoken dialogue. We use pyannote.audio \cite{Plaquet23, Bredin23} to calculate statistics for the relevant turn-taking events. For semantic evaluation, we employ the Whisper-v3 speech recognition model to transcribe spoken dialogues into text. We then concatenate the utterances from different speakers, inserting a speaker turn token $<st>$ between them. Finally, we use the DialoGPT model \cite{zhang2019dialogpt} to calculate the perplexity of the text transcripts of the audio continuations. The generation temperature is always set to 1.

\subsection{Results and Analysis}

\begin{figure*}[t]
    \centering
    \includegraphics[width=\textwidth]{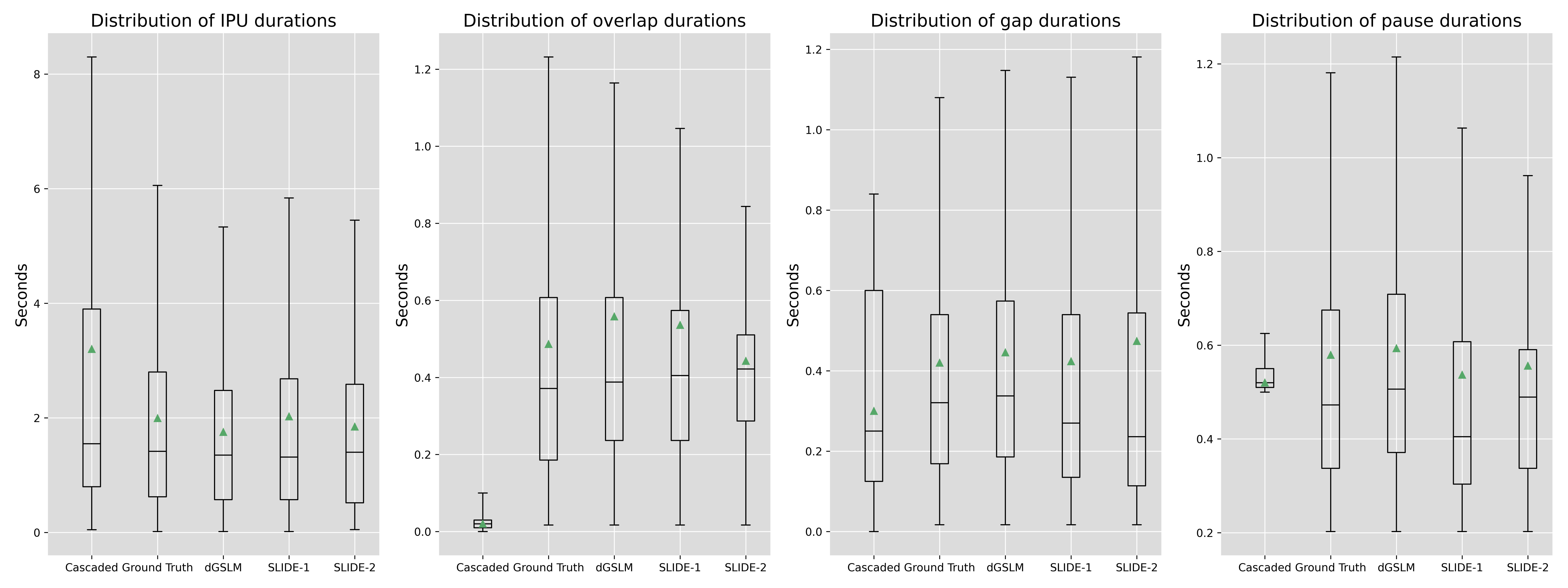}
    \caption{The temporal distribution of turn-taking events. The green triangles denote the mean values, and the solid lines within the boxes represent the median.}
    \label{fig:box}
\end{figure*}

\subsubsection{Objective Evaluations on Spoken Dialogue Naturalism}

We calculate the temporal distribution statistics for four types of turn-taking events: IPUs, overlaps, gaps, and pauses. We report the number of occurrences per minute and the cumulated durations of each turn-taking event per minute in Tables \ref{tab:occurrences_per_min} and \ref{tab:durations_per_min}, respectively. In addition, we present the overall temporal distribution of these turn-taking events using box plots in Fig. 2.

In Tables \ref{tab:occurrences_per_min}, \ref{tab:durations_per_min} and \ref{tab:ppl&mos}, the results for the Cascaded model are cited from \cite{dgslm}, while we reproduce dGSLM and calculate the statistics. The notation SLIDE-1 means that the textual dialogue is directly copied from the test dataset. SLIDE-2 means that the textual dialogue is generated by LLMs.

The results in Tables \ref{tab:occurrences_per_min} and \ref{tab:durations_per_min} indicate that while there are some slight differences in the turn-taking patterns of spoken dialogues generated by SLIDE and dGSLM, they significantly outperform the cascaded model.

Fig. \ref{fig:box} illustrates the box plots of the temporal distribution of turn-taking events across different models. The results indicate that, with the exception of the cascaded system, SLIDE and dGSLM exhibit turn-taking event statistics that closely resemble those of the ground truth spoken dialogue. We can conclude that the performance of SLIDE is comparable to that of the original dGSLM in terms of naturalism in spoken dialogue.

\begin{table}[b]
    \centering
    \renewcommand\arraystretch{1.5}
    \caption{Perplexity and MOS of the Generated Spoken Dialogues and Ground Truth Spoken Dialogues}
    \fontsize{9pt}{11pt}\selectfont 
    \begin{tabular}{l|ccc}
        \hline
        Model & Perplexity $\downarrow$ & N-MOS $\uparrow$ & M-MOS $\uparrow$ \\
        \hline\hline
        Cascaded \cite{dgslm} & - & 2.38 $\pm$ 0.63 & 2.70 $\pm$ 0.38 \\
        dGSLM & 1228.82 & 4.14 $\pm$ 0.78 & 1.52 $\pm$ 0.40 \\
        SLIDE-1 & 532.81 & \textbf{4.37} $\pm$ \textbf{0.46} & 3.94 $\pm$ 0.81 \\
        SLIDE-2 & \textbf{421.29} & 4.06 $\pm$ 0.41 & \textbf{4.08} $\pm$ \textbf{0.49} \\
        \hline
        Ground Truth & 371.16 & $4.72 \pm 0.40$ & $4.63 \pm 0.44$ \\
        \hline
    \end{tabular}
    
    \label{tab:ppl&mos}
\end{table}

\subsubsection{Objective Evaluations on Spoken Dialogue Semantic Coherence}

We conduct experiments based on the generated textual dialogues or the transcripts of the ground truth spoken dialogues. For each model, we report the perplexity of the first 50 words in the transcripts of the generated audio continuations.

    
    

From Table \ref{tab:ppl&mos}, we can conclude that our proposed method significantly improves semantic coherence compared to the original dGSLM, with a relative perplexity reduction of 65.8\%, from 1228.82 to 421.29. Notably, our method achieves performance comparable to the ground truth spoken dialogue. The perplexity of the ground truth spoken dialogue is 371.16, while SLIDE-2 reaches a perplexity of 421.29, representing a relative disparity of only 11.9\%.

\subsubsection{Subjective Evaluations}

We ask the participants to rate N-MOS (Naturalness-MOS) and M-MOS (Meaningfulness-MOS). The specific rating rules have been illustrated in Section \ref{subsec: metric}, and the MOS results are shown in Table \ref{tab:ppl&mos}.

In Table \ref{tab:ppl&mos}, SLIDE-1 achieves the highest N-MOS of 4.37 $\pm$ 0.46, slightly outperforming the original dGSLM. SLIDE-2 achieves the highest M-MOS of 4.08 $\pm$ 0.49, marking a significant improvement over the original dGSLM, with a relative improvement of 270.0\%. Compared to the ground truth M-MOS of 4.63 $\pm$ 0.44, SLIDE-2 has a relative disparity of only 11.9\%. We can conclude that the proposed method significantly enhances the meaningfulness of the generated spoken dialogues compared to the original dGSLM.

\section{Conclusion}
\label{sec:conclusion}

In this paper, we propose the SLIDE model for spontaneous spoken dialogue generation that combines the strengths of both traditional cascaded systems and SLM-based systems. By leveraging a hybrid scheme that utilizes text to capture semantic context and speech units to preserve paralinguistic information, our method addresses key challenges in generating naturalistic and semantically coherent dialogues. Through extensive experiments and evaluations, we demonstrated that our proposed method significantly improves the semantic coherence of generated dialogues while maintaining the naturalistic characteristics. Our results show that the perplexity of the generated spoken dialogue has been reduced from 1228.82 to 421.29, and also approaches to the performance of the ground-truth spoken dialogues.



\bibliographystyle{IEEEtran}
\bibliography{refs}

\end{document}